\documentclass[aps,pra,preprint,showpacs,groupedaddress]{revtex4}
\usepackage{graphicx} 
\usepackage{amsmath}

\begin{document}

\title{Transverse excitations of ultracold matter waves upon \\
propagation past abrupt waveguide changes}
\author{M.~Koehler}
\author{M.W.J.~Bromley}
  \email{bromley@phys.ksu.edu}
\author{B.D.~Esry}
  \email{esry@phys.ksu.edu}
\affiliation{Department of Physics, Kansas State University, Manhattan, KS 66506 USA}

\date{\today}

\vspace{1cm}

\begin{abstract}

The propagation of ultracold atomic gases through abruptly changing
waveguide potentials is examined in the limit of non-interacting atoms.
Time-independent scattering calculations of microstructured waveguides
with discontinuous changes in the transverse harmonic binding potentials
are used to mimic waveguide perturbations and imperfections.
Three basic configurations are examined: step-like, barrier-like and well-like
with waves incident in the ground mode.  At low energies, the spectra rapidly
depart from single-moded, with significant transmission and reflection of
excited modes.  The high-energy limit sees $100\%$ transmission,
with the distribution of the transmitted modes determined simply by
the overlap of the mode wave functions and interference.

\vspace{.7cm}

\end{abstract}

\pacs{03.75.Be 03.75.Kk 03.65.Nk}

\maketitle

\vspace{1cm}

\section{Introduction}

The manipulation of ultracold matter waves can now, somewhat routinely,
be performed above microchip or magnetized surfaces \cite{hinds99a,folman02a}.
In such experiments, the quantum nature of the dilute atomic gases dominates
over the classical, enabling precision matter wave control \cite{bongs04a}.

One of the key requirements in using an ``atom chip'' to perform atom optics
is the ability to transport atoms from one atom optical component to another.
Here, we present calculations of wave propagation
through waveguides with idealized perturbations consisting
of sudden changes to the transverse confining potential. 
An increase (decrease) in the tranverse confining potential
results in a decrease (increase) in the kinetic energy along
the direction of wave propagation, providing effective step
potentials along the waveguide.

The present study was motivated in three ways.  Firstly, recent
experiments have demonstrated a significant fragmentation
of a Bose-Einstein condensate (BEC) in a waveguide located close to
the surfaces \cite{leanhardt02a,fortagh02b}, attributed, at
least in part, to imperfections created during the wire
fabrication processes \cite{kraft02a,leanhardt03a,wang04a,esteve04a,vale04a}.
Secondly, to further understand some of the limitations to
designing atom optics devices that are based on variations
of the waveguide potentials, for example, the smoothly varying
wide--narrow--wide wire geometry has been proposed as a
quantum-point-contact type device for atoms \cite{thywissen99b}.
Thirdly, the literature has been lacking a multimode analysis of
many of the simplest waveguide geometries, as it has been
experimentally shown that introducing a perturbation in a waveguide
can result in the transverse excitation of a BEC \cite{leanhardt02a}.

To characterize the impact of transverse discontinuities, here
we explore the Schr{\"o}dinger wave mechanics of waveguides with
step-like, barrier-like and well-like potentials along the direction
of propagation.  There have already been some theoretical investigations
using time-dependent calculations of wave propagation through smooth
potentials such as a bottleneck (step-up) and a smooth termination
(extreme step-down) in the limit of non-interacting atoms
\cite{jaaskelainen02a,jaaskelainen02c}, while non-linear (atom-atom) effects
in the bottleneck-type geometry have also been examined \cite{stickney02a,lahaye03a}.
The advantages in using abrupt potentials whilst neglecting
atom-atom interactions is that simple time-independent calculations can
be used to characterize the transmission and reflection probabilities.
Under these conditions, we have previously investigated a
circular bend \cite{bromley03b}, which consists of an abrupt
transition from the lead waveguides into the bend and at
low-energies behaves like a potential well.

Our multimode analysis, restricted as it is to the the linear
regime, provides a baseline for comparison of
BEC propagation through quasi-one-dimensional (1-D) waveguides
including the transverse degrees of freedom.
For example, previous time-independent studies have 
investigated non-linear wave propagation through
shallow-well, step and gaussian shaped 1-D potentials
\cite{leboeuf01a,pavloff02a,leboeuf03a,seaman04a}.
Such simple waveguide potentials could be generated by modifying
the transverse confinement, where knowledge of the transverse
excitation probabilities, in the abrupt and linear limits, should be useful.

For ground mode matter waves propagating at low energies through
the various perturbations, the present results show that the spectra
rapidly depart from single-moded, with significant transmission and
reflection of excited modes.  The high-energy limit sees
$100\%$ transmission, and we present a simple model to determine the
distribution of the transmitted modes that combines the overlap of the
mode wave functions with the multi-path interference of the modes.

\section{Details of the calculations}

There are a number of atom chip wire configurations that
can create waveguides \cite{thywissen99a,folman02a}, but we follow
the theoretical ansatz adopted in Refs.~\cite{bromley03b,bromley04a}.
That is, we assume that the waveguides consist of an idealized trapping
potential that is quadratic near the minimum and operate at low matter-wave
densities such that atom-atom interactions can be neglected.
Furthermore, so that the waveguide potentials reduce to an effective
2D problem, the waveguides are assumed to be created by multiple
wire configurations with abrupt changes in the spacing between the wires,
such that the height and transverse position of the potential minima
remains constant along the waveguide.  The out-of-plane quantum number
is then a conserved quantity.

We consider simple harmonic oscillator (SHO) confining potentials,
which, for barrier-like or well-like effective potentials, are given by
\begin{equation}
\label{eqn:pots2D}
V(x,z) = \begin{cases}
      \frac{1}{2} m \omega_a^2 x^2, & \quad z \le z_0 \; , \\
      \frac{1}{2} m \omega_b^2 x^2, & \quad z_0 \le z \le z_1\; , \\
      \frac{1}{2} m \omega_a^2 x^2, & \quad z \ge z_1 \; .
\end{cases}
\end{equation}
The barrier-like potential has $\omega_b>\omega_a$; the
well-like potential has $\omega_a>\omega_b$.
The step-like potential consists of only one change in frequency.

\begin{figure}[th]
\includegraphics[width=3.5in]{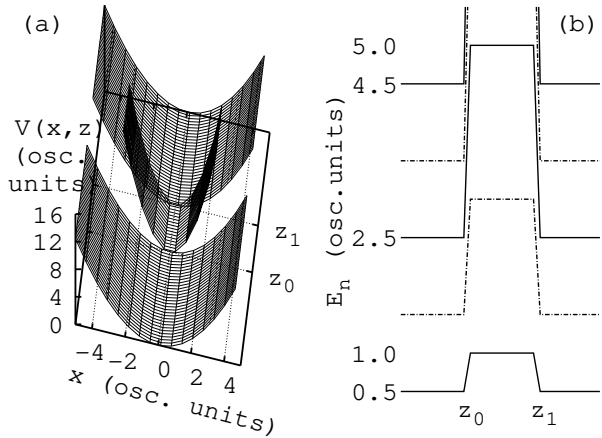}
\caption[]{ \label{fig:abruptpot}
(a) Potential energy surface of a barrier-like waveguide with
$\omega_a = 1$ and $\omega_b = 2$ and barrier length $l = z_1 - z_0$.
(b) Energy levels of the leads and barrier transverse SHO potential
along $z$.  The solid lines at $z_0$ and $z_1$ should be vertical,
but instead are drawn on an angle to highlight the lead--barrier--lead transition.
The dot-dashed lines correspond to the parity-forbidden levels
(assuming an even incoming mode).
All the energies and coordinates are given in terms of
oscillator units for the leads.
}
\end{figure}

Oscillator units are used throughout this paper, where energies
are in units of $\hbar \omega$, while lengths are given in units
of $\beta = \sqrt{\hbar / m\omega}$.  An example barrier-like
potential is shown in Fig.~\ref{fig:abruptpot}(a), where the
reference frequency is $\omega_a = 1$ and $\omega_b = 2$.

The corresponding energy levels of Eq.~(\ref{eqn:pots2D}) are shown
in Fig.~\ref{fig:abruptpot}(b).  These energy levels
behave as effective potentials for the longitudinal motion since
we expand the total wavefunction in each region
on the transverse oscillator states.
In this model, all coupling between modes occurs through
the matching between regions.  The potentials in
Eq.~(\ref{eqn:pots2D}) are symmetric in $x$ so that parity in $x$
is conserved, simplifying the present analysis and discussion
considerably [the dot-dashed lines in Fig.~\ref{fig:abruptpot}(b)
are not coupled to the solid lines].   In experiments, imperfections
would as likely be off-center, resulting in populating of all
modes (as was possible in our previous study of the
circular bend \cite{bromley03b}).  The fundamental physics, however,
remains much the same, so we chose to adopt parity-conserving
perturbations with the incoming waves restricted to the
ground (even) oscillator mode.

To perform the time-independent scattering calculations we
initially adopted the transfer matrix method \cite{merzbacher70a},
although most of the calculations reported in this paper use the
interface matching method \cite{merzbacher70a,bromley03b}.  The two
methods are similar, however, and since the transfer matrix approach
facilitates the discussion of our results, we outline it here.
The extension of the transfer matrix method from
one-dimension to include transverse degrees of freedom is
trivial \cite{wu93b,pereyra02a}, so only a short summary is given
here as it applies to a barrier/well-like geometry of length $(z_1-z_0)=l$.

Firstly, the time-independent wavefunction is expanded on transverse SHO states,
$\varphi_n(x)$ for frequencies $\omega_a$, and $\chi_m(x)$ for $\omega_b$:
\begin{equation}
\label{eqn:wavefunk2D}
\begin{split}
\Psi_\mathrm{I}(x,z) &= \sum_n \varphi_n(x) \;
           \big[a_n e^{i k_n (z-z_0)} + b_n e^{-i k_n (z-z_0)}\big] \quad z \le z_0 \; ,\\
\Psi_\mathrm{II}(x,z) &= \sum_m \chi_m(x) \;
           \big[c_m e^{i \kappa_m (z-z_0)} + d_m e^{-i \kappa_m (z-z_0)}\big] \quad z_0 \le z \le z_1 \; , \\
\Psi_\mathrm{III}(x,z) &= \sum_n \varphi_n(x) \;
           \big[g_n e^{i k_n (z-z_1)} + h_n e^{-i k_n (z-z_1)}\big] \quad z_1 \le z \; .
\end{split}
\end{equation}
where the momenta are $k_n = \sqrt{2E-\omega_a(2n+1)}$ and
$\kappa_m = \sqrt{2E-\omega_b(2m+1)}$.  Matching the wavefunctions
and their first derivatives across each interface and then projecting
out the modes gives the following sets of equations:
\begin{equation}
\label{eqn:matrices}
\begin{pmatrix}
\vec{a} \\
\vec{b}
\end{pmatrix}
=
\begin{pmatrix}
A^{11} & A^{12} \\
A^{21} & A^{22}
\end{pmatrix}
\begin{pmatrix}
\vec{c} \\
\vec{d}
\end{pmatrix}
\quad \textrm{and} \quad
\begin{pmatrix}
\vec{c} \\
\vec{d}
\end{pmatrix}
=
\begin{pmatrix}
B^{11} & B^{12} \\
B^{21} & B^{22}
\end{pmatrix}
\begin{pmatrix}
\vec{g} \\
\vec{h}
\end{pmatrix} \; .
\end{equation}
The matrix elements of each submatrix are
\begin{equation}
\label{eqn:mes}
\begin{split}
A^{11}_{nm} &= A^{22}_{nm} = \frac{1}{2}\Big(1+\frac{\kappa_m}{k_n}\Big) O_{nm} \; ,  \\
A^{12}_{nm} &= A^{21}_{nm} = \frac{1}{2}\Big(1-\frac{\kappa_m}{k_n}\Big) O_{nm} \; , \\
B^{11}_{mn} &= e^{-2i\kappa_m l} B^{22}_{mn} = \frac{1}{2}\Big(1+\frac{\kappa_m}{k_n}\Big)e^{-i\kappa_m l} O_{nm} \; , \\
B^{12}_{mn} &= e^{-2i\kappa_m l} B^{21}_{mn} = \frac{1}{2}\Big(1-\frac{\kappa_m}{k_n}\Big)e^{-i\kappa_m l} O_{nm} \; ,
\end{split}
\end{equation}
using the notation $O_{nm} = \int \varphi_n(x) \chi_m(x)\:dx$.
While generating functions are known for the overlaps of SHO-functions with
different frequencies \cite{birtwistle77a,aslangul95a} (such overlaps are also
found in calculations of transitions between molecular vibrational
modes \cite{manneback51a}), we performed the transverse integrations
numerically using a B-spline basis.

The scattering solution is obtained by constructing the transfer matrix $Q = AB$,
which relates one lead's coefficients, $\vec{a}$ and $\vec{b}$, to the other's,
$\vec{g}$ and $\vec{h}$.  Given that the wave is restricted to
incoming from $z<z_0$ in the ground mode ($a_0 = 1$, $a_{n_i>0} = 0$,
and $\vec{h} = \vec{0}$), the linear equation $\vec{a} = Q^{11} \vec{g}$ is
solved, and then $\vec{b}= Q^{21}\vec{g}$.
The transmission and reflection probabilities for each mode are then given by:
\begin{equation}
\label{eqn:transref}
\mathrm{T}_{n_f} = \frac{|g_{n_f}|^2 k_{n_f}}{|a_0|^2 k_0}
\quad \textrm{and} \quad
\mathrm{R}_{n_f} = \frac{|b_{n_f}|^2 k_{n_f}}{|a_0|^2 k_0} \; .
\end {equation}

There are convergence difficulties with the transfer matrix approach,
the demonstration and discussion of which is mostly relegated to
the appendix.  In brief, the problems are related to the slow decay of
the SHO overlaps $O_{0m}$ with $m$, which requires the
inclusion of strongly closed channels, leading to the appearance of large
exponentials in the transfer matrix.  While the transfer matrix method
eliminates the need to find intermediate coefficients
(ie. $\vec{c}$ and $\vec{d}$), it was generally found
to be unstable beyond the smallest of the perturbations considered in
this paper.  The interface matching method, in contrast, explicitly
solves for the intermediate coefficients and is able to include
enough closed channels to ensure near machine-precision convergence
for the range of geometries and energies given in this paper.
The transfer matrix method works well, however, for step-like geometries
with a single interface (ie. $Q=A$) as there are no
exponentials in the matrix elements, and furthermore, all the
coefficients are explicitly solved for.

\section{Results}

The multimoded transmission and reflection probabilities for three
basic geometries are given in this section: the step-like potential,
the barrier-like potential and the well-like potential.
The calculations for the step-like potentials use the transfer matrix
method, while interface matching is used for the barrier-like
and the well-like potentials.   Before presenting these calculations,
it is instructive to discuss how the interface overlaps $O_{0m}$
scale with frequency.

\subsection{Interface overlaps}

The $O_{0m}$ dependence on the frequency ratio $\omega_b/\omega_a$
is discussed here since it strongly influences the amount of
mode excitation caused by the different geometries.
Using the orthonormality of the SHO eigenstates, along with the
recursion relations of the Hermite polynomials, it can be shown
that the overlap integrals of the $\omega_a=1$ ground mode with
the $\omega_b$ even modes reduce to the particularly simple form:
\begin{equation}
\label{eqn:overlaps}
O_{0m} = \sqrt{\frac{2\sqrt{\omega_b}\;(m-1)!!}{(\omega_b+1)\;m!!}}\;
             \Big(\frac{\omega_b-1}{\omega_b+1}\Big)^{m/2} \;,
\end{equation}
for all $m=0,2,4,6\hdots$.  Due to symmetry, $O_{0m}=0$ for $m=1,3,5\hdots$.
Eq.~(\ref{eqn:overlaps}) was also obtained by Aslangul \cite{aslangul95a}.

\begin{figure}[th]
\includegraphics[width=3.5in]{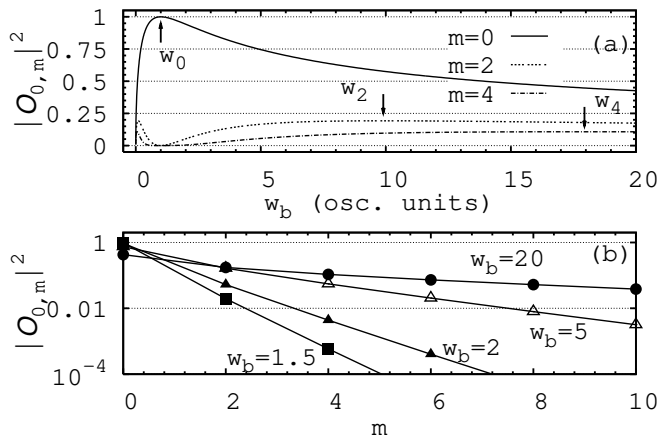}
\caption[]{ \label{fig:omegadependence}
SHO wavefunction overlaps $|O_{0m}|^2$ between the ground mode
with fixed frequency $\omega_a=1$, and another mode with
variable frequency $\omega_b$.
(a) Shows the $m=0,2,4$ modes as a function of $\omega_b$.
The arrows labelled by $\omega_m$ indicate the maxima.
(b) Shows the overlaps as a function of $m$ for four
frequencies: $\omega_b=1.5$ (squares),  $\omega_b=2$ (filled-triangle),
$\omega_b=5$ (hollow-triangle) and  $\omega_b=20$ (circles).
The lines between dots in (b) are added to guide the eye.
The frequencies shown in both (a) and (b) are given in
oscillator units relative to $\omega_a$.
}
\end{figure}

The dependence of the first three overlaps ($m=0,2,4$) on frequency
is shown in Fig.~\ref{fig:omegadependence}(a).
At $\omega_b=1$, there is a perfect waveguide match, and
we must have $O_{00} = 1$ and $O_{0m} = 0$ for $m>0$.
As $\omega_b$ increases, $O_{00}$ monotonically decreases
towards zero.  At the same time, the overlap with
each excited mode increases and reaches a maximum
when the characteristic equation,
$\omega_b^2-2(2m+1)\omega_b+1=0$, is satisfied.
As a function of frequency, these maxima occur at
$\omega_m=(2m+1)+\sqrt{(2m+1)^2-1}$, which, for $m=2$, is 
$\omega_{2}=9.8989$ and for $m=4$, is $\omega_{4}=17.944$.
Comparing the width ($\sqrt{\langle x^2 \rangle}$)
of the SHO functions at these frequencies against the
ground mode's reveals that
$\sqrt{\langle x_m^2(\omega_m) \rangle} \approx \sqrt{1/4} = \sqrt{\langle x_0^2 \rangle/2}$.
This connection, while natural, is not particularly
illuminating and is not pursued here any further.
Past the maxima, the overlaps slowly decrease as $\omega_b^{-1/4}$
to zero.  Due to symmetry of the ratio $\omega_b/\omega_a$,
a second maxima of $|O_{0m}|^2$ also exists at $1/\omega_m$.

The slow decay of the overlaps with the higher modes
can be seen in Fig.~\ref{fig:omegadependence}(b).
This implies that, even at low energies, many closed channels
must be included in the following calculations to achieve
computational convergence.

\subsection{Step-like waveguide potential}

\begin{figure}[th]
\includegraphics[width=3.5in]{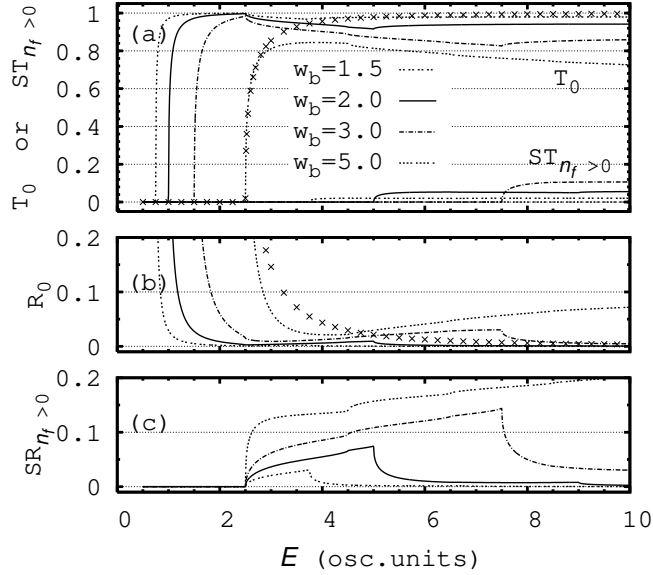}
\caption[]{ \label{fig:stepwithenergy}
Multimoded scattering probabilities of step-like potentials
which consist of a single abrupt change in the transverse
confinement potential from $\omega_a=1$ to $\omega_b=1.5$,$2$,$3$, and $5$.
The incoming waves are in the ground state $n_i=0$.
(a) Shows $\mathrm{T}_{0}$ and $\sum \mathrm{T}_{n_f>0}$, the
transmission probabilities into the ground mode and the sum
of the transmission probabilities into the excited modes,
respectively.
(b) Gives $\mathrm{R}_{0}$, the reflection probabilities into
the ground mode, while (c) gives $\sum \mathrm{R}_{n_f>0}$, the
sum of the reflection probabilities into the excited modes.
The total energy $E$ is given in oscillator units
relative to $\omega_a$.
The crosses in (a) and (b) are the analytic transmission
and reflection probabilities for a 1-D step potential for the case
of $\omega_b=5$ (ie. of height $V_0=|2.5 - 0.5|$ with a background
potential $V=0.5$ added to correct the reflection ground mode threshold).
}
\end{figure}

The transmission and reflection probabilities, $\mathrm{T}_{n_f}$
and $\mathrm{R}_{n_f}$, of ground mode plane-waves traversing four
step-like waveguides are shown in Fig.~\ref{fig:stepwithenergy}. 
At incident energies below the lowest reflection threshold
($E < 2.5$, the lowest excited mode energy), the system behaves like
the familiar 1-D step potential.  Ground state transmission, $\mathrm{T}_{0}$,
remains the dominant channel across the range of energies
shown in Fig.~\ref{fig:stepwithenergy}(a)
(for the range of $\omega_b$ examined here),
although excited mode transmission can also be seen
in Fig.~\ref{fig:stepwithenergy}(a) as each mode opens. 
Significant reflection is seen into the ground $n_f=0$ mode
in Fig.~\ref{fig:stepwithenergy}(b), which rapidly
drops off from threshold.  As each reflection threshold
opens, the reflection into the excited modes $n_f>0$,
seen in Fig.~\ref{fig:stepwithenergy}(c), firstly
increases then is seen to experience an overall decrease.
All of which are consistent with the
Wigner threshold laws for multichannel systems \cite{fano86a}.

For incidence in the ground channel, the total transmission
approaches $100\%$ in the high energy limit.  In this limit
the transmission probability is given simply by
\begin{equation}
\label{eqn:project}
\mathrm{T}_{n_f}(E\to\infty) = |O_{0n_f}|^2 \; ,
\end{equation}
with the SHO overlaps of Eq.~(\ref{eqn:overlaps}).
Such projections were introduced as part of the waveguide
calculations of J{\"a}{\"a}skel{\"a}inen and Stenholm
\cite{jaaskelainen02a,jaaskelainen04a}, in which
the transmission excitation probabilities generated by both
bottleneck and split-potential waveguides were briefly discussed
as the potentials tended towards abrupt.  Similar multimode
projections have also been theoretically examined during
expansion of a BEC from a microtrap into a waveguide \cite{stickney02a},
with an emphasis on the effects of atom-atom interactions.

Equation~(\ref{eqn:project}) can be seen as the high-energy limit
of the matrix elements of $A$, given by Eq.~(\ref{eqn:mes}).
At energies high compared to the step height, the momenta of the
lowest few modes are approximately the same on both sides of the
step, $\kappa_{n_f} \approx k_0$.  Given that the overlaps
limit the number of channels involved, the matrix elements that
dominate the $Q=A$ transfer matrix are then
$A^{11}_{0,n_f} = A^{22}_{0,n_f} \approx O_{0n_f}$,
while $A^{12}_{0,n_f} = A^{21}_{0,n_f} \approx 0$.
For an incoming wave in the ground mode with
$a_0 = 1$, $a_{n_i>0} = 0$, and $\vec{d}=0$,
then the outgoing waves have $c_{n_f} \approx O_{0n_f}$ while $\vec{b}=0$.

\begin{figure}[th]
\includegraphics[width=3.5in]{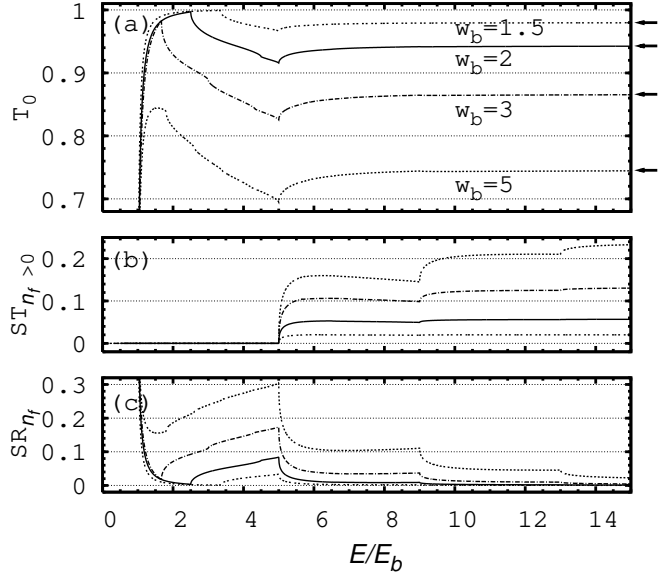}
\caption[]{ \label{fig:stepwithenergyscaled}
The scattering probabilities of the same four step-like potentials
as Fig.~\ref{fig:stepwithenergy} ($\omega_b=1.5$,$2$,$3$, and $5$)
plotted here as a function of $E/E_b$, where $E_{b} = \omega_b/2$.
For each waveguide, (a) shows the transmission probabilities
of the ground mode, (b) the sum of the excited mode
transmission probabilities $\sum \mathrm{T}_{n_f>0}$,
and (c) the total reflection probabilities $\sum \mathrm{R}_{n_f}$.
The arrows at $E/E_b=15$ correspond to the $|O_{00}|^2$ interface
overlaps.
}
\end{figure}

To more clearly show this limit, the results of
Fig.~\ref{fig:stepwithenergy} are replotted in
Fig.~\ref{fig:stepwithenergyscaled} on an energy axis scaled by
$\omega_b$ instead of $\omega_a$ so that the transmission
channels for each waveguide open at the same scaled energy.
At high energies, the transmission probabilities shown
in Fig.~\ref{fig:stepwithenergyscaled}(a) increase towards
asymptotes of $\mathrm{T}_{0}(E\to\infty) = |O_{00}|^2$,
in agreement with Eq.~(\ref{eqn:project}).
For the four different waveguides shown here, the limits
are $\mathrm{T}_{0}(E\to\infty) = 0.979796$, $0.942809$, $0.866026$,
and $0.745356$.
Figures~\ref{fig:stepwithenergyscaled}(b) and (c) show that
at the onset of each transmitted mode (ie. at $E/E_b = 5, 9$ etc.)
the transmission probability into that mode increases, taking flux
from reflection.
The mismatch in mode wavefunctions for $\omega_b=5$, for example,
is particularly severe, with
$\mathrm{T}_{n_f=0,2,4,6}(E\to\infty) = 0.745356$, $0.165635$,
$0.055212$, and $0.020449$ [see Fig.~\ref{fig:omegadependence}(b)].
In this case, these four transmission modes must be open before the
high energy limit [$\sum \mathrm{T}_{n_f} = 1$, as per Eq.~(\ref{eqn:project})]
is reached to within $2\%$.

Whether the wave is incident from the left or the right,
these results apply.  The ground mode
transmission $\mathrm{T}_{0}(E)$ is \textit{absolutely identical}
as a function of total energy $E$ for both the waveguide
constriction (ie. step-up from $\omega_a=1$
to $\omega_b>1$), and the waveguide expansion (ie. step-down
from $\omega_a>1$ to $\omega_b=1$)
This results also holds for the familiar 1-D step potential, and, although
we do not show it here, the transmission and reflection mode mixing
conspires to ensure this is also the case in the multichannel system.
The mode mixing as a function of $E$ is different for either geometry,
however, since for the step-up case there can be many reflection
channels open at the lowest transmission threshold, $E=E_b$, while
for the step-down, there can be many transmission channels open at $E=E_b$.
At high energies, neither of the step geometries generates
reflection, and for incidence in the ground mode the limits
from Eq.~(\ref{eqn:project}) apply.

\subsection{Barrier-like waveguide potential}

To demonstrate the characteristics of a single barrier-like potential,
we consider the case shown in Fig.~\ref{fig:abruptpot},
for a fixed length $l$ and
frequencies that change from $\omega_a=1$ to $\omega_b>\omega_a$
and back to $\omega_a=1$.  We also present
the high-energy characteristics for scattering from this potential.

\begin{figure}[th]
\includegraphics[width=3.5in]{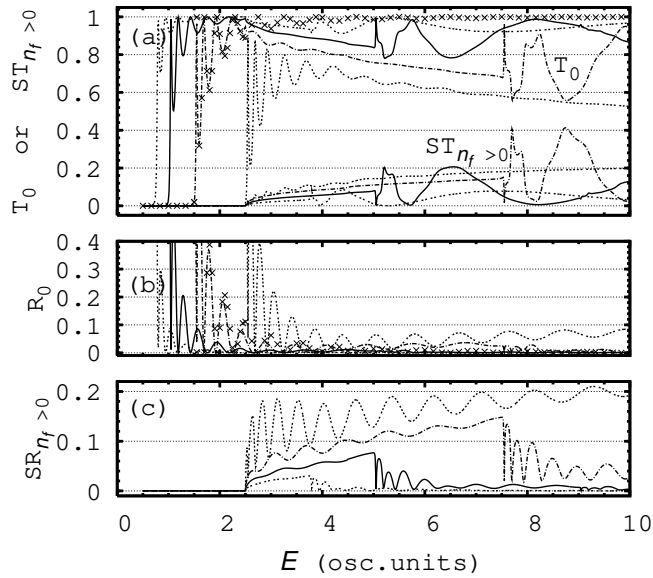}
\caption[]{ \label{fig:barrierwithenergy}
Multimoded scattering probabilities of four barrier-like potentials
with $\omega_b=1.5$,$2$,$3$, and $5$ for a fixed length $l=10$ as a function
of $E$ (in oscillator units relative to $\omega_a$). 
The legend for the different frequencies is the
same as Figs.~(\ref{fig:stepwithenergy}) and (\ref{fig:stepwithenergyscaled}).
The organisation of the probabilities is also the
same as Fig.~(\ref{fig:stepwithenergy}).
The crosses in (a) and (b) are the analytic transmission
and reflection probabilities for a 1-D barrier potential for the case
of $\omega_b=3$ (ie. of height $V_0=|1.5 - 0.5|$ with a background
potential $V=0.5$).
}
\end{figure}

The transmission and reflection probabilities for four waveguide
constrictions $\omega_b=1.5$,$2$,$3$,$5$ and length $l=10$ are shown
in Fig.~\ref{fig:barrierwithenergy}.
Resonances appear in all of the spectra at low energies due to
the wavelength matching condition $n\lambda/2 \approx l$.
The 1-D analytic results for $\omega_b=3$ in
Fig.~\ref{fig:barrierwithenergy}(a) highlight the transmission
resonances as the result of low-energy ground-mode propagation
over the barrier.
As soon as the $\omega_a=1$, $n=2$ mode opens at $E=2.5$, however,
multichannel physics takes over ($n=1$ is not allowed due to symmetry).
Above $E=2.5$, there is significant excited mode
transmission [Fig.~\ref{fig:barrierwithenergy}(a)],
while the amount of reflection into the ground mode
[Fig.~\ref{fig:barrierwithenergy}(b)] and the excited modes
[Fig.~\ref{fig:barrierwithenergy}(c)] is significant across
the energy range.

\begin{figure}[th]
\includegraphics[width=3.5in]{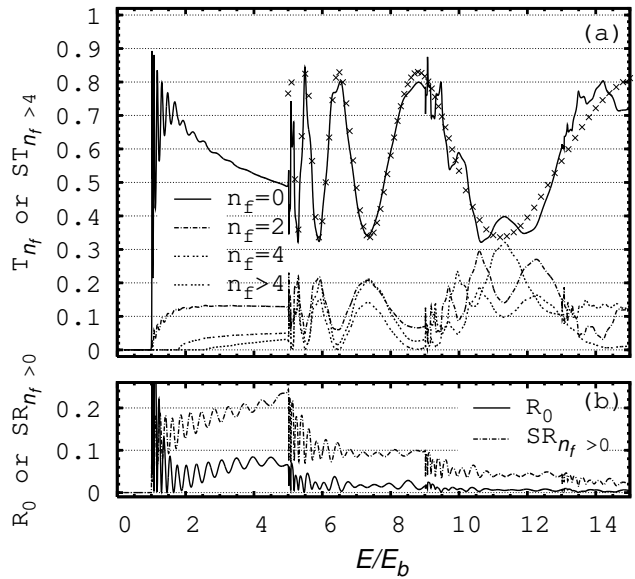}
\caption[]{ \label{fig:barrierwithenergyscaled}
The scattering probabilities of the $\omega_b=5$, $l=10$, barrier-like
potential seen in Fig.~\ref{fig:barrierwithenergy} plotted here
as a function of $E/E_b$ where $E_b=2.5$.
(a) Shows the individual transmission probabilities of the $n_f=0,2,4$ modes
alongside the sum of the transmission probabilities for the $n_f>4$ modes.
(b) Shows the ground-mode reflection probabilities as well as
the sum of the excited mode reflection probabilities.
The crosses in (a) correspond to the $\mathrm{T}_{0}(E\to\infty)$
two-mode interference model of Eq.~(\ref{eqn:interf}).
The first peak of $T_0$ reaches nearly up to 1, which is not shown
due to the limited energy resolution of the figure.
}
\end{figure}

To observe the high-energy limit, the results for
the $\omega_b=5$ barrier of Fig.~\ref{fig:barrierwithenergy}
were extended to higher energies, and are shown in
Fig.~\ref{fig:barrierwithenergyscaled} with the energy scaled by $E_b=2.5$.
Figure~\ref{fig:barrierwithenergyscaled}(b) shows that
while there is more structure in the reflection probabilities
than for the step-like potential
in Fig.~\ref{fig:stepwithenergyscaled}(c), the total amount
of reflection still tends towards zero as more barrier modes
become open at $E/E_b=5,9,13,\hdots$. 

To obtain an expression analogous to Eq.~(\ref{eqn:project}),
we must take into account the fact that the transmitted waves
going through a barrier experience at least two interface
projections as in Eq.~(\ref{eqn:project}).
The phases accumulated while propagating the length $l$ of the barrier
must also be included in such a prescription, which suggests that at
high energies (relative to the barrier height)
\begin{equation}
\label{eqn:interf}
\mathrm{T}_{0}(E\to\infty) = \Big|\sum_{m} e^{i(\kappa_m-\kappa_0)l} O_{0m}O_{m0}\Big|^2 \; ,
\end{equation}
where $m$ is only summed over the propagating barrier modes.
The crosses shown in Fig.~\ref{fig:barrierwithenergyscaled}(a)
demonstrate that the two-mode version of this model does a remarkable job
in describing the transmission probability above the $m=2$ barrier
threshold (at $E/E_{b}=5$).  Depending on the phase differences, the modes that
are excited at the first interface can be converted back to the ground
mode by the second interface.  We also noted this behavior in
circular waveguide bends \cite{bromley03b}, where the amount of
excitation could be suppressed by changing the angle swept out
by the bend to the point where the accumulated phase difference
between the $n=0$ and $n=1$ modes was a multiple of $\pi$.
A similar design consideration could perhaps be useful for
atom optics devices such as the quantum point contacts \cite{thywissen99b},
where any unavoidable -- yet unwanted -- mode excitations could be
minimised by varying the length between the changes in waveguide potentials.

\subsection{Well-like waveguide potential}

The transmission and reflection probabilities for well-like waveguides
due to a potential bulge ($\omega_a > \omega_b$)
are the focus in this last section.  The scattering behavior for a well
is complicated by the presence of bound states which translate into
the presence of Feshbach resonances in a multichannel problem.
Much of the resonance physics seen here has been extensively discussed as part
of our studies of the circular waveguide bend \cite{bromley03b,bromley04b}.
The propagation thresholds in a bend lie slightly lower than
the connecting leads \cite{exner89a,goldstone92a}, resulting in very weakly
bound states and energetically narrow resonances.  The present
well-like waveguides can provide extreme differences between the lead
and bulge energy thresholds, and thus the possibility of multiple
narrow resonances located below the thresholds.

\begin{figure}[th]
\includegraphics[width=3.5in]{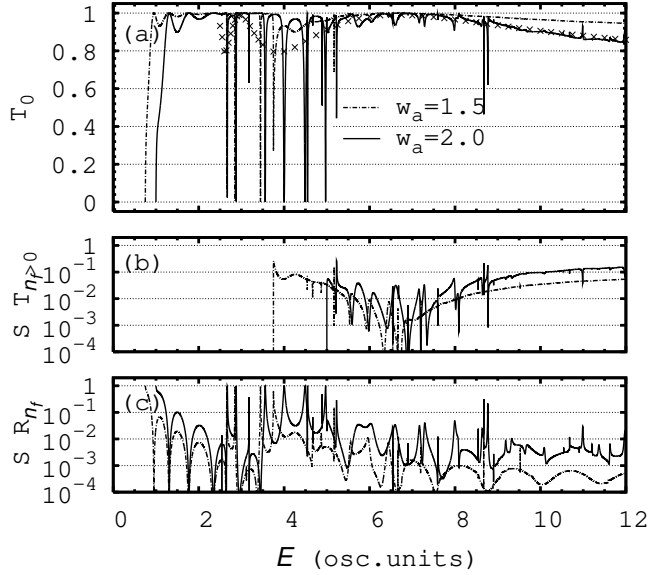}
\caption[]{ \label{fig:wellwithenergy}
Multimoded scattering probabilities of two well-like potentials
with $\omega_a=1.5$ and $2$ for a fixed length $l=10$ as a
function of $E$ (in oscillator units relative to $\omega_b$).  The
organisation of the probabilities is the same
as Fig.~\ref{fig:stepwithenergyscaled}.
The crosses in (a) correspond to the $\mathrm{T}_{0}(E\to\infty)$ two-mode
interference model of Eq.~(\ref{eqn:interf}) for $\omega_a=2$.
}
\end{figure}

Two bulges are considered here, the first from $\omega_a=1.5$
to $\omega_b=1$ and back to $\omega_a=1.5$, and the second with
$\omega_a=2$ to $\omega_b=1$ and back to $\omega_a=2$.  In both
cases, we choose $\omega_b=1$ to be the reference oscillator frequency,
to simplify the comparison with the barriers in the previous section.
The transmission and reflection probabilities of both of these
potentials with well length $l=10$ are shown in
Fig.~\ref{fig:wellwithenergy}.

Multiple $100\%$ reflection resonances exist at energies below the first
excited lead mode energy, ie. below $E=3.75$
for $\omega_a=1.5$ and $E=5$ for $\omega_a=2$.  As was noted in our
circular bend studies \cite{bromley03b}, the reflection resonances below
the second excited mode (ie. below $E=6.75$ for $\omega_a=1.5$, and
at $E=9$ for $\omega_a=2$) do not result in complete reflection due
to the reduced coupling between the ground and second excited mode
and due to the existence of alternate pathways to reflection.

The high-energy transmission probability asymptotes for the well-like
potential are again given by Eq.~(\ref{eqn:interf}), although the
energy of the first excited mode within the well means that the
two-mode model starts at $E=2.5$ (in other words, at a lower energy than
for the equivalent barrier-like potential).  The two-mode model is shown
in Fig.~\ref{fig:wellwithenergy}(a) for the $\omega_a=2$ well,
and is seen to be a bad approximation near the $E=2.5$ threshold
due to the significant reflection there.
The two-mode model generally provides a reasonable approximation
at higher energies once the total reflection probability has dropped
below the total excited mode transmission probability, and also when
the many narrow reflection resonances no longer play a role.

Waveguides with more extreme discontinuities such as $\omega_a=3$
and $5$ were also explored.  They exhibit so many resonances across
the range of energies shown in Fig.~\ref{fig:wellwithenergy},
however, that the transmission and reflection probabilities
essentially become a dense series of closely-spaced
vertical spikes.  At energies above the second excited
mode threshold ($E=4.5$), there are not so many resonances,
although there remains significant reflection probability. 
As an example, the $\omega_a=5$, $l=10$, well-like waveguide has a
total reflection probability at $E=37.5$ ($E/E_a=15$)
that has only dropped down to about $0.2$.  This reflection probability can be
compared with the $\omega_b=5$, $l=10$ barrier-like geometry in
Fig.~\ref{fig:barrierwithenergyscaled}(b) at $E=37.5$ ($E/E_b=15$),
where the total reflection probability was only about $0.02$.
In other words, reflections play a far more dominant role for
the well-like potentials than the barrier-like.

\section{Summary}

Using time-independent solutions of the Schr{\"o}dinger equation,
we have explored the propagation of dilute, ultracold atomic gases
through abruptly changing waveguide potentials.
Previous studies have discussed the conditions for ``adiabatic''
waveguide propagation through microstructures 
(eg. Refs.~\cite{thywissen99b,jaaskelainen02a,jaaskelainen02c,bortolotti04a}).
In contrast, the interest here was on the consequences of sudden potential
variations for mode excitation with a view towards modelling waveguide imperfections,
examining the effects of using abrupt potentials in atom optical devices,
and simply to explore the behavior of some simple geometries.

Three idealized geometries with changes in the transverse guiding potential
have been the focus of study: step-like, barrier-like and well-like.
The low-energy behavior of all the geometries departed from
single-moded, with the exception of the mildest perturbations
at energies below the lowest excitation threshold.
Significant generation of both transmission and reflection excited modes
was caused by the mismatch of the modes at the interfaces between the
waveguide sections.  The strong coupling to excited modes is due to the
significant overlap of the ground SHO function of one frequency with
the excited SHO functions with a different frequency.  Certainly, care
should be taken during wire fabrication of atom optical elements to ensure
that any deliberate (or not) changes in a waveguides transverse frequency
are not abrupt.

High energy wave propagation through abrupt potentials amounts
to $100\%$ wave transmission via projections across each interface,
along with multimode interference.
The present SHO-based waveguides behave somewhat differently
than the hard-walled models for ballistic electron propagation
through waveguides with a wide-narrow junction
(see Ref.~\cite{szafer89a}, and the references thereafter
that cite it).  In that case an impedence mismatch occurs,
where there is always some amount of reflection at high energies due
to the fact that the narrow guide modes can never represent the wide
guide modes over the whole width of the lead.  Although, to the best of
our knowledge, calculations for such electron waveguides have not discussed
high-energy/abrupt potential transmission and reflection limits.

A further condition was suggested for high energy transmission
through microstructures with multiple interfaces to account for the
interference between modes.  It was shown that a simple two-mode
model can give a reasonable approximation to the amount of
ground mode transmission, and provides an additional consideration
for the design of ``atom-chip'' waveguides to control single-moded
wave-transmission through potentials that generate multimoded excitations.

\begin{acknowledgments}
This research was supported by the Department of the
Navy, Office of Naval Research, and also by the Research Corporation.
\end{acknowledgments}

\appendix

\section{Convergence of transfer matrix calculations}
\label{sec:convergence}

Our transfer matrix program was validated by reproducing the single-mode
calculations of electron propagation through a linear array of
1-D potentials \cite{wu91a}, and secondly by comparing with multi-moded
2-D results from an interface matching program \cite{bromley03b}.

The multimode transfer matrix method, however, has numerical convergence
difficulties as the number of modes included in a calculation increases.
This behavior can be seen in Table \ref{tab:transconv}, for scattering
with different incident energies $E$ off a barrier-like potential with
$\omega_b/\omega_a = 2$ of length $l=5$.  The unitarity of the transfer
matrix results deviate significantly from
$\sum_f \mathrm{T}_{n_f} + \mathrm{R}_{n_f} = 1$, even as $N$ is increased.
The interface matching calculations are shown for comparison, and barely
suffer from the same problems.
\begingroup
\squeezetable
\begin{table*}[th]
\caption[]{
\label{tab:transconv}
The convergence of transfer matrix and interface matching calculations
for a barrier-like potential of length $l=5$ with $\omega_b/\omega_a = 2$
The number of channels included in each calculation is given by $N$.
The ground mode transmission probabilities and unitarity
(given by $1-\sum_f (\mathrm{T}_{n_f} + \mathrm{R}_{n_f})$)
are given for three energies: $E=0.75$, where only barrier tunnelling
is energetically allowed, and $E=6$, where three lead modes and two
barrier modes are open.
}
\begin{ruledtabular}
\begin{tabular}{lcccc}
  & \multicolumn{2}{c}{$E=0.75$} & \multicolumn{2}{c}{$E=6$} \\
$N$ & $\mathrm{T}_0$ & Unitarity & $\mathrm{T}_0$ & Unitarity \\
\hline
\hline
 \multicolumn{5}{c}{Analytic (1-D barrier)} \\
1  & 0.00339154 & --- & 0.999976 & --- \\
\hline
 \multicolumn{5}{c}{Transfer Matrix $Q=AB$} \\
1  & 0.00381548 & --4.2394$^{-4}$ & 1.124973 & --1.2500$^{-1}$ \\
2  & 0.00267029 &   3.0328$^{-5}$ & 0.772832 & --5.2616$^{-2}$ \\
4  & 0.00274164 &   2.8303$^{-7}$ & 0.804769 &   8.1844$^{-4}$ \\
8  & 0.00274223 &   4.1060$^{-7}$ & 0.807601 & --3.4533$^{-8}$ \\
16 & 0.00264476 & --9.4352$^{-3}$ & 0.849707 & --3.4386$^{-2}$ \\
\hline
 \multicolumn{5}{c}{Transfer Matrix $Q=AC^{-1}$} \\
1  & 0.00339154 &   3.3307$^{-16}$ & 0.999976 &   2.0912$^{-16}$ \\
2  & 0.00270062 & --1.4129$^{-12}$ & 0.795176 &   3.3826$^{-2}$ \\
4  & 0.00274192 &   1.7813$^{-10}$ & 0.805169 & --3.1198$^{-3}$ \\
8  & 0.00274223 & --4.5390$^{-6}$  & 0.807546 & --1.5052$^{-5}$ \\
16 & 0.00277009 &   4.0555$^{-3}$  & 0.916579 & --5.6275$^{-2}$ \\
\hline
 \multicolumn{5}{c}{Interface Matching} \\
1  & 0.00339154 & 2.2204$^{-16}$ & 0.999976 &   0 \\
2  & 0.00270062 & 1.1102$^{-16}$ & 0.795176 &   3.3826$^{-2}$ \\
4  & 0.00274192 & 3.3307$^{-16}$ & 0.805169 & --3.1198$^{-3}$ \\
8  & 0.00274223 & 0              & 0.807605 & --3.8357$^{-6}$ \\
16 & 0.00274223 & 4.4409$^{-16}$ & 0.807607 & --7.6739$^{-13}$ \\
32 & 0.00274223 & 1.1102$^{-16}$ & 0.807607 &   2.2205$^{-16}$ \\
\end{tabular}
\end{ruledtabular}
\end{table*}
\endgroup

In an attempt to rescue the transfer matrix approach, a second method for constructing
the transfer matrix was investigated.  This second attempt involved inverting a matrix $C$,
\begin{equation}
\label{eqn:matrices2}
\begin{pmatrix}
\vec{c} \\
\vec{d}
\end{pmatrix}
=
\begin{pmatrix}
C^{11} & C^{12} \\
C^{21} & C^{22}
\end{pmatrix}^{-1}
\begin{pmatrix}
\vec{g} \\
\vec{h}
\end{pmatrix} \; ,
\end{equation}
with matrix elements,
$C^{11}_{nm} = C^{22}_{nm} = \frac{1}{2}(1+\kappa_m/k_n) O_{nm}$,
and $C^{12}_{nm} = C^{21}_{nm} = \frac{1}{2}(1-\kappa_m/k_n) O_{nm}$
while the matrix elements in $A$ of Eq.~(\ref{eqn:mes}) were modified to include the
$e^{\pm i\kappa_ml}$ terms.  The numerical results of solving $Q=AC^{-1}$ are also
shown in Table \ref{tab:transconv}, and while unitarity results for the
single-channel calculation, it is as unstable, if not worse that $Q=AB$.

For the geometries considered in this paper the transfer matrix approach
was not able to adequately converge over the required range of energies.
The exception to this was for the step-like geometry, where for the most
extreme calculations in this paper
(Fig.~\ref{fig:stepwithenergyscaled}: $\omega_b/\omega_a=5$ and $E=37.5$)
$N=128$ modes could easily be included (since the transfer matrix contains no
exponential terms, and all coefficients are explicitly solved for).

We also find that, since the $O_{nm}$ overlaps are energy independent,
obtaining convergence is much easier for the present interface matching
calculations than for previous circular waveguide bend calculations
using the same method \cite{lin96a,bromley03b}.
For the most extreme barrier-like calculation
(Fig.~\ref{fig:barrierwithenergyscaled}: $\omega_b/\omega_a=5$, $l=10$ and $E=37.5$),
the interface matching method achieved unitarity to better
than $10^{-10}$ with $N=96$ modes.

Some of the transfer matrix convergence issues were passingly mentioned
by Wu \textit{et.al} \cite{wu93b}, where the transfer matrix method
was initially applied to a system of multiple circular bends
(the transfer matrix is simply built as a number of matrix products).
In the end, however, the calculations shown in Ref.~\cite{wu93b} were
performed using multiple interface matching.  
We labor on this point since no such unitarity problems were
noted in the multichannel transfer matrix calculations of Ref.~\cite{pereyra02a},
where the examples given were for 2 or 3 channel calculations with
delta potentials.  On the other hand, the advantage of the transfer
matrix approach is that, in the single-mode approximation, one can easily
obtain the transmission characteristics of both finite and infinite length
periodic potentials \cite{merzbacher70a,bromley04c}.

\end{document}